# Beyond Synthesis Conditions: Linking *In-situ* Growth Dynamics to Carbon Nanotube Forest Properties through Multimodal Machine Learning

Shun Muroga[*ab†], Takashi Tsuji[a†], Shunsuke Sakurai[a], Don N Futaba[*ab]

[a]: Nanocarbon Material Research Institute, National Institute of Advanced Industrial Science and Technology (AIST), Tsukuba Central 5, 1-1-1, Higashi, Tsukuba, Ibaraki, 305-8565, Japan

[b]: Materials DX Research Center, National Institute of Advanced Industrial Science and Technology (AIST), Tsukuba Central 5, 1-1-1, Higashi, Tsukuba, Ibaraki, 305-8565, Japan

[†] contributed equally

* Corresponding author. S. Muroga (muroga-sh@aist.go.jp), D. N. Futaba (d-futaba@aist.go.jp)

**Abstract**

Machine learning for chemical vapor deposition (CVD) of carbon nanotubes (CNTs) has largely mapped static synthesis conditions to final properties, implicitly treating these conditions as a complete description of growth. Here, we test this assumption by integrating gas-flow conditions with *in-situ* time-resolve growth profiles (growth curves) of vertically aligned CNT forests. We introduce an observation-window analysis, in which prediction is evaluated as a function of observation time by providing models with the growth curve only up to progressively later times, to determine not only whether a property can be predicted, but when the necessary information becomes available during synthesis. Clustering of normalized growth curves revealed distinct growth patterns independent of final forest height. The gas conditions alone could predict the initial growth rate but not the growth lifetime and failed to capture the variation in yield ($R^2 < 0$). Adding the growth curve improved yield prediction ($R^2 \approx 0.4$–$0.5$), with the relevant information emerging during the intermediate and later stages of growth rather than at its onset. In contrast, density and G/D ratio, which forest height does not record, showed no consistent improvement. The maximum height and lifetime could be estimated with moderate accuracy midway through the 10-min growth. Permutation feature importance identified water vapor and $H_2$, both linked to catalyst activity, as the most informative gas variables. These results indicate that variation unexplained by synthesis conditions is not merely noise but reflects an evolving catalyst state, and they establish mid-growth decision points for autonomous synthesis, enabling early termination of unpromising batches and, ultimately, in-batch adjustment of growth conditions.

## 1. Introduction

Carbon nanotubes (CNTs) are one-dimensional, tubular nanomaterials composed of carbon atoms and have attracted considerable attention since their discovery because of their unique and exceptional physical and chemical properties [1,2]. Their high strength, thermal conductivity, and electron mobility, combined with their low density, have motivated applications ranging from composites with polymers, rubbers, and metals to components for batteries and semiconductor devices. Realizing these applications, however, depends on synthesizing CNTs with controlled yield and structure.

Chemical vapor deposition (CVD) is one of the most widely used methods for CNT synthesis because it enables efficient and scalable CNT growth and structural control. Extensive studies, ranging from fundamental research to industrial-scale production, have focused on controlling CNT properties such as yield, crystallinity, and diameter, and on elucidating the underlying growth mechanisms [3–8]. One notable advance is the water-assisted (super-growth) CVD method [2], in which the addition of a small amount of water maintains catalyst activity and enables the highly efficient synthesis of high-purity vertically aligned CNTs (VACNT) forests. In this process, the forest height, and therefore the yield, depends not only on the initial growth rate but also on how long the catalyst remains active, which is described as the growth lifetime [3,4].

Controlling CNT growth nevertheless remains a challenge because CVD is a complex and dynamic process in which multiple interacting factors jointly determine the outcome. These factors include catalyst stability, catalyst activity and lifetime, carbon-source supply, and the flow of various process gases. Importantly, catalyst activity is not fixed by the synthesis conditions, but evolves during the growth, so the final properties reflect the history of the process as well as the inputs. To capture this history, considerable effort has been devoted to the direct *in-situ* observation of CNT growth. For example, Puretzky et al. employed laser- and image-based techniques to characterize CNT growth [9], Abdulhafez et al. applied video image processing for growth analysis [10], and Dee et al. [11] reported an electron-microscopy-based approach. These studies revealed time-dependent growth behavior, including changes in growth rate and growth termination. However, such measurements have mainly been used to characterize growth kinetics after synthesis, rather than to anticipate the growth termination.

Recent advances in data science and machine learning have enabled the extraction of correlations from complex, multidimensional experimental data and their use for materials-property prediction, interpretation of underlying phenomena, and optimization of process conditions [12–17]. Combined with automated experimentation, these methods have accelerated the development of self-driving laboratories [18,19] capable of autonomously exploring experimental space. In the field of CNTs, data-driven approaches have been applied to various material forms, including dispersions [17], films [12,16,20], composites [14], fibers [15], and increasingly to synthesis itself [21–26].

For example, Nikolaev, Maruyama, and co-workers developed the Autonomous Research System

(ARES), which integrated *in-situ* Raman spectroscopy with machine learning [21]. Subsequent studies reported Bayesian optimization of CNT growth rates by Chang et al. [22], exploration of conditions for diameter-selective CNT growth by Rao et al. [23], and analysis of catalyst activity and multi-objective Bayesian optimization by Waelder et al. [27] Data-driven prediction has also been applied to other CNT synthesis processes, including the prediction of CNT yield in aerosol CVD [25], the relationship between CNT yield and diameter in catalytic methane decomposition [26], and the average CNT diameter predicted from literature [28]. For CNT forests, we previously optimized growth efficiency and crystallinity [13], while Li et al. used large language models for catalyst discovery [29]. Other studies have combined simulation with machine learning, such as molecular dynamics simulations based on machine-learning interatomic potentials to investigate CNT nucleation, defect formation, and defect-healing [30], and finite element simulations combined with physics-informed neural networks to analyze CNT growth and mechanical deformation [31].

Despite this progress, most of these studies share a common formulation: synthesis parameters such as catalyst composition, gas composition and flow rates, and temperature are mapped directly to final properties such as yield, height, crystallinity, or diameter. This formulation implicitly treats the synthesis conditions as a complete description of the growth. Even when *in-situ* signals are acquired, as in ARES [21], they typically serve to evaluate the outcome of each experiment rather than as inputs that reveal the state of the process. Two questions central to understanding and controlling CNT growth therefore remain unanswered. First, how much of the final outcome is set by the synthesis conditions, and how much emerges only during growth? Second, at what point during a single growth does sufficient information become available to anticipate its outcome? The second question is particularly important for autonomous synthesis, because it determines whether a growth process can be evaluated, and potentially redirected, before the process finishes.

In this study, we address these questions by integrating gas-flow conditions with *in-situ* growth curves of VACNT forests in a machine-learning framework. We first establish what the gas conditions alone can predict, for both the final forest properties and the growth kinetics. We then introduce an observation-window analysis, in which models receive the growth curve only up to progressively later times, to determine not only whether each property can be predicted, but when the necessary information becomes available. Finally, we use permutation feature importance to identify which stages of growth and which process gases contribute most to yield prediction. This approach provides a basis for understanding CNT synthesis in terms of an evolving process state and for self-driving laboratories capable of real-time process-state recognition and in-batch decision-making.

## 2. Experimental

### 2.1 CNT Synthesis

$Al_2O_3$/Fe (40 nm/1.8 nm) underlayer/catalyst films were prepared by sequential sputtering on 20 mm × 20 mm Si wafers with a 500 nm thermally grown $SiO_2$ layer. The formation of catalyst nanoparticles was carried out by a fast-heating process using a pre-heated 1-inch quartz tube furnace maintained at 750 °C, followed by annealing in $H_2$ ambient for 6 min to transform the Fe catalyst film into nanoparticles.

Subsequently, CNT forests were synthesized by water-assisted chemical vapor deposition at 750 °C for 10 min under atmospheric pressure. Acetylene ($C_2H_2$) was used as the carbon source, water vapor as the growth enhancer, and $H_2$ and $N_2$ as carrier gases. In contrast to conventional water-assisted CVD processes, $H_2$ was intentionally introduced during the forest growth stage and treated as an experimental parameter. The total gas flow rate was fixed at 600 sccm throughout all experiments, while the growth temperature (750 °C) and growth time (10 min) were also kept constant. Three process parameters were varied: $H_2$ flow rate (0-480 sccm), the $C_2H_2$ flow rate (10-30 sccm), and the $N_2$ flow rate supplied through a water bubbler (50-400 sccm), which determined the amount of water vapor introduced into the reactor. Acetylene was supplied from a gas cylinder containing 20 vol% $C_2H_2$ diluted in $N_2$. The balance of the total gas flow was adjusted with $N_2$ so that the overall flow rate remained constant at 600 sccm.

The synthesized VACNT forests were characterized in terms of areal yield (mg $cm^{-2}$), forest height ($\mu$m), CNT forest density (mg $cm^{-3}$), and Raman G/D ratio. The CNT forest density was calculated from the measured yield and forest height. *In-situ* height monitoring [4] was performed during growth to obtain forest growth curves, from which the initial growth rate, maximum height, and growth lifetime were extracted. These parameters were used as response variables in the machine-learning analysis.

### 2.2 Machine Learning Analysis

Machine-learning regression was performed to investigate the relationships among synthesis conditions, *in-situ* CNT growth trajectories, and the resulting CNT properties. Three input configurations were considered: (i) gas-flow conditions only, (ii) *in-situ* height profiles only, and (iii) a combination of gas-flow conditions and height profiles. The target variables included the yield, CNT forest density, G/D ratio, initial grow rate, maximum height, and growth lifetime.

The *in-situ* height profiles were interpolated onto a common time grid comprising 200 points. Because neighboring time points were strongly correlated, principal component analysis (PCA) was incorporated into the machine-learning pipeline for models using height-profile information. The PCA transformation was fitted exclusively to the training data within each cross-validation fold to prevent information leakage. The number of retained principal components was selected from 2, 4, 6, and 8 within the inner cross-validation procedure.

Six regression algorithms were compared: ridge regression, LASSO regression, partial least-squares regression (PLS), Gaussian process regression (GPR), random forest regression, and XGBoost regression. GPR employed a Matérn kernel with $\nu$ = 1.5 combined with a white-noise kernel. Hyperparameters for the applicable regression algorithms were optimized within the inner cross-validation loop.

Model development and evaluation were performed using nested cross-validation. Five-fold cross-validation constituted the outer loop for estimating generalization performance, while the inner cross-validation loop was used for hyperparameter and PCA-dimension selection based on the mean absolute error (MAE). Predictions for each sample were obtained from an outer-fold model for which that sample was not included in model training, providing out-of-fold (OOF) predictions for the complete dataset. Predictive performance was evaluated using the coefficient of determination ($R^2$), MAE, and root mean squared error (RMSE). The same cross-validation framework was applied to all input configurations and regression algorithms to enable a consistent comparison of predictive performance.

To investigate when predictive information emerged during CNT growth, cumulative observation-window analysis was performed using height profiles acquired from the beginning of growth to 2, 4, 6, 8, and 10 min. A separate regression model was constructed for each observation window using the same nested cross-validation framework. Gas-only models were used as the 0-min baseline when evaluating the combined gas-flow and height-profile models. This analysis quantified the progressive gain in predictive information as the in-situ observation period increased.

In addition, permutation feature importance was evaluated over all measured time points of the height profile. The increase in MAE following random permutation of each feature was used as its importance measure, and the calculation was repeated ten times to obtain the mean and standard deviation. For the combined gas-flow and height-profile models, the importance of the three gas-flow variables was evaluated on the same scale as the time-resolved height-profile importance. This analysis was used to identify the temporal regions utilized by the predictive models rather than to infer direct causal effects on CNT growth.

## 3. Results and Discussion

### 3.1 Experimental framework and *in-situ* growth monitoring

To examine whether direct observation of vertically aligned carbon nanotube (VACNT) forest growth provides information beyond the synthesis conditions, we varied the gas-flow conditions while maintaining a constant growth temperature, synthesis duration, and total gas flow (Experimental). Each experiment combined controlled gas-flow parameters, time-resolved growth profiles, and measurements of the resulting forest properties (height, mass, density, and intensity ratio of the Raman graphitic to disorder peaks (G/D ratio)). These complementary data allowed us to evaluate whether the observed growth response provides additional information for predicting the final forest properties and at what stage during the growth that information becomes available.

CNT forests were synthesized by chemical vapor deposition (CVD) on catalyst-coated substrates (Fig. 1a). The as-grown sample exhibited a macroscopically continuous layer of CNTs extending from the substrate surface (Fig. 1b), and cross-sectional scanning electron microscopy (SEM) confirmed the densely packed, vertically aligned bundled structure (Fig. 1c).

The evolution of the forest height was monitored during the synthesis to obtain growth curves, which capture the overall progression of the forest growth from initial stages through subsequent growth and, in some cases, termination (Fig. 1d). Forest height and growth kinetics, such as growth lifetime and initial growth rate, were extracted from these data [3,4]. The time-dependent height measurements were also used as input features for machine-learning analysis of the relationships among synthesis conditions, growth behavior, and resulting CNT properties.

We first examine how the final forest properties vary with gas-flow conditions as a starting point for evaluating their predictability.

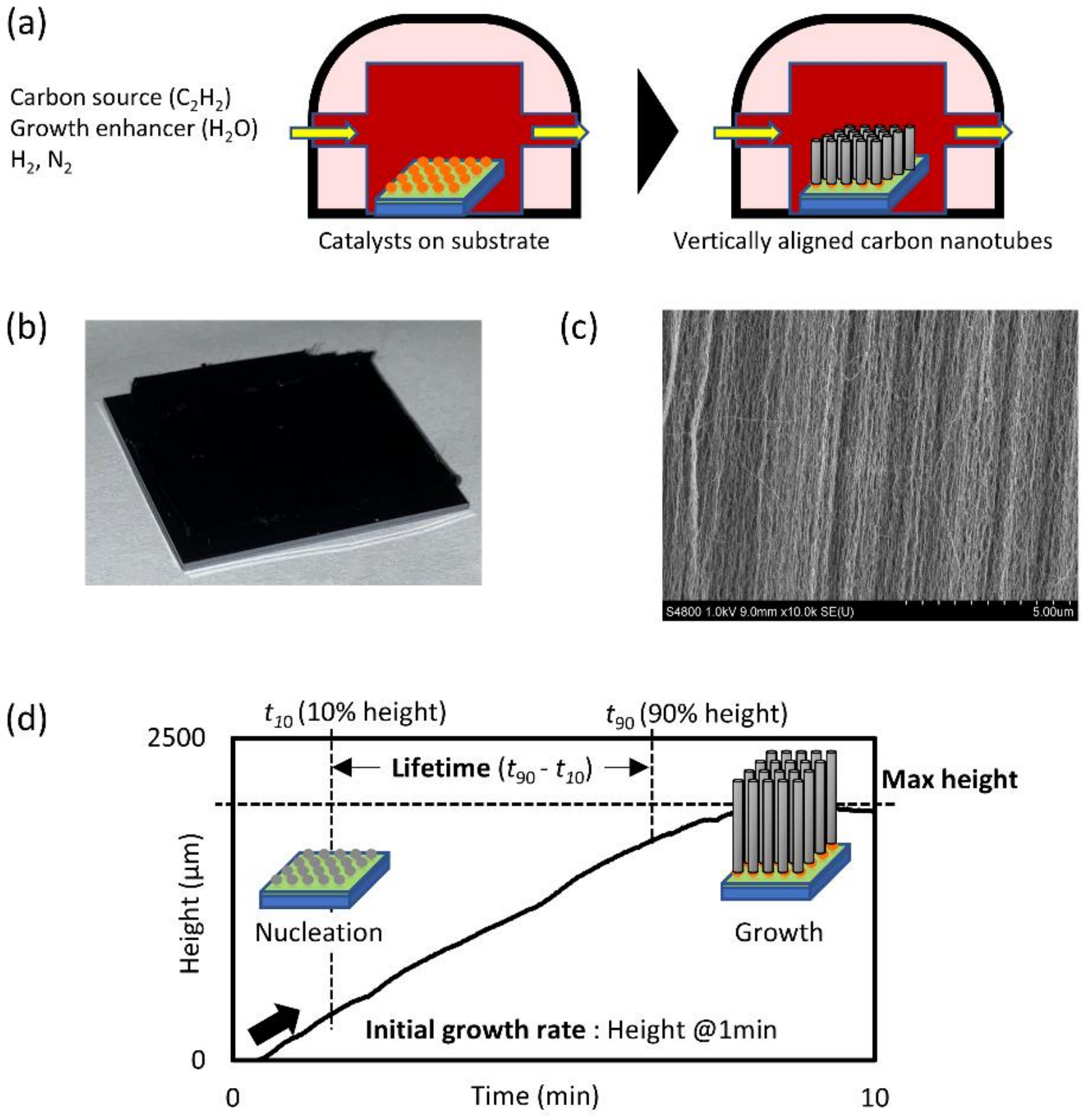

**Fig. 1.** Overview of CNT synthesis, morphology, and growth monitoring. (a) Schematic illustration of the chemical vapor deposition (CVD) process for the synthesis of vertically aligned carbon nanotube (VACNT) forests. (b) Representative photograph of an as-grown VACNT forest. (c) Representative cross-sectional scanning electron microscopy (SEM) image of the VACNT forest. (d) Schematic

illustration of the temporal evolution of VACNT forest growth and the definition of the growth lifetime.

### 3.2 Relationship between gas conditions and VACNT properties

Because the value of growth observations can only be judged against what the synthesis inputs already explain, we first examined how the final forest properties depend on the gas-flow conditions. The resulting VACNT properties, including yield, CNT density, and G/D ratio, varied substantially across the investigated gas-flow conditions (Figure 2). This observation is consistent with previous reports showing that synthesis conditions strongly influence both the amount and quality of the resulting CNT forests.

For CNT yield (Fig. 2a), substantial variations were observed even at identical individual gas flow rates, with no clear monotonic dependence on the $H_2O$, $C_2H_2$, or $H_2$ flow rate. Although relatively high yields were observed under several specific conditions, the considerable scatter indicates that no single gas-flow variable alone can adequately explain the variation in CNT yield. CNT density (Fig. 2b) showed broadly similar variations with the gas-flow conditions, while considerable scatter was again observed at the same or similar individual gas flow rates. The G/D ratio (Fig. 2c), an indicator of the structural quality of the CNTs, also varied considerably across the investigated conditions.

Notably, none of these VACNT properties exhibited a simple monotonic dependence on any single gas flow rate. Samples synthesized at the same value of an individual gas flow rate exhibited substantially different yields, CNT densities, and G/D ratios. These results indicate that individual gas-flow variables alone are insufficient to describe the variation in the final forest properties and that the combined effects of multiple synthesis parameters need to be considered. Therefore, describing and predicting the resulting VACNT properties requires consideration of the multidimensional synthesis conditions.

While these results demonstrate the dependence and substantial variability of the synthesis outcomes with respect to the externally controlled gas conditions, they do not describe how the CNT forests actually evolved during growth. The *in-situ* growth curves (Fig. 1d), in contrast, record the temporal evolution of individual CNT forests throughout the growth process. Before evaluating whether this time-resolved information improves the prediction of the final forest properties, we therefore examined whether the growth profiles differ only in the attained forest height or also in their temporal evolution.

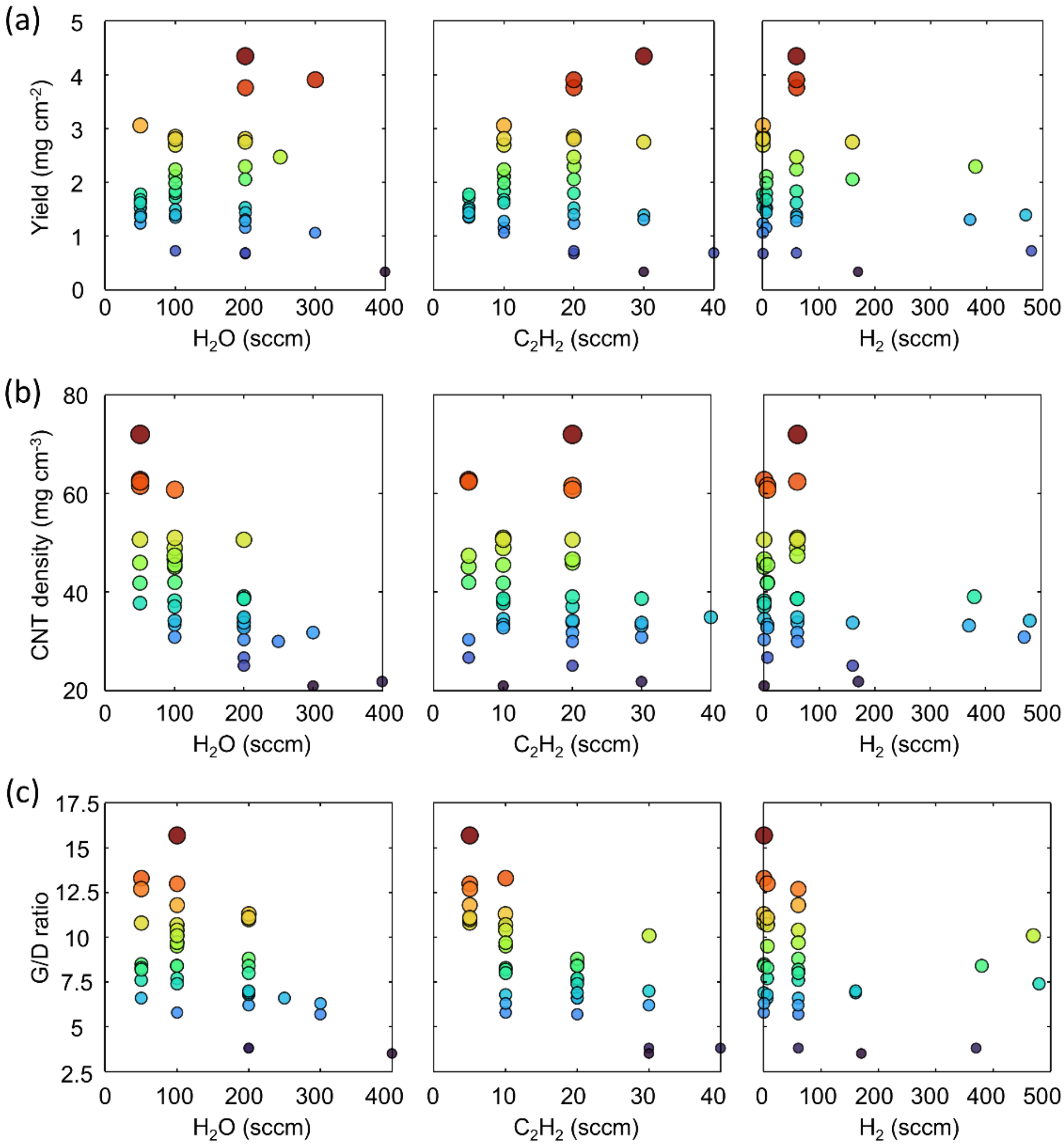

**Fig. 2.** Relationships between gas flow conditions and the resulting VACNT forest properties: (a) CNT yield, (b) CNT density, and (c) G/D ratio. The relationships with the water content, carbon feedstock content, and $H_2$ content are shown from left to right in each row.

### 3.3 Analysis of the *in-situ* growth curves

If the growth curves differed only in the attained forest height, they would provide little more than a final-height measurement. However, the growth curves differed in shape as well as in magnitude (height range), and distinct temporal growth patterns persisted even after normalization to the final height of the growth period (Fig. 3). These differences in shape reflected the initial growth rates, growth persistence, and whether growth terminated within the 10-min growth period. We identified these patterns using principal component analysis (PCA) and hierarchical clustering of both the raw and normalized growth curves.

Hierarchical clustering of the raw profiles identified five representative growth patterns that differed not only in the final forest height but also in growth rate and the onset of growth termination (Fig. 3a). The corresponding PCA score plot showed a clear separation among these clusters, with the first principal component (PC1) accounting for 92.8% of the total variance (Fig. 3b). The dominance of a single component suggests that the variation among the raw profiles is largely governed by differences in overall growth magnitude, which could mask differences in the curve shape.

To separate shape from magnitude, we analyzed normalized growth curves to distinguish differences in the temporal growth behavior from those in growth magnitude. After normalization, characteristic differences in the shape remained evident (Fig. 3c). For example, the representative profiles exhibited different temporal behaviors, ranging from relatively gradual growth over the observation period to rapid growth followed by an early termination. Hierarchical clustering of the normalized curves identified four representative growth patterns. In the corresponding PCA score plot (Fig. 3d), PC1 and PC2 accounted for 83.5% and 13.1% of the variance, respectively (96.6% in total). Unlike the raw curves, in which a single component dominated, the normalized curves required a second component to describe their variation, and forests of similar final height were assigned to different clusters.

Thus, the *in-situ* growth curves provide a dynamic descriptor of the CVD process that the final forest height alone does not capture. Because the final properties also varied among samples grown at the same or similar gas flow conditions (Fig. 2), these differences in growth behavior may help account for that variability. We therefore first establish what the gas conditions alone can explain.

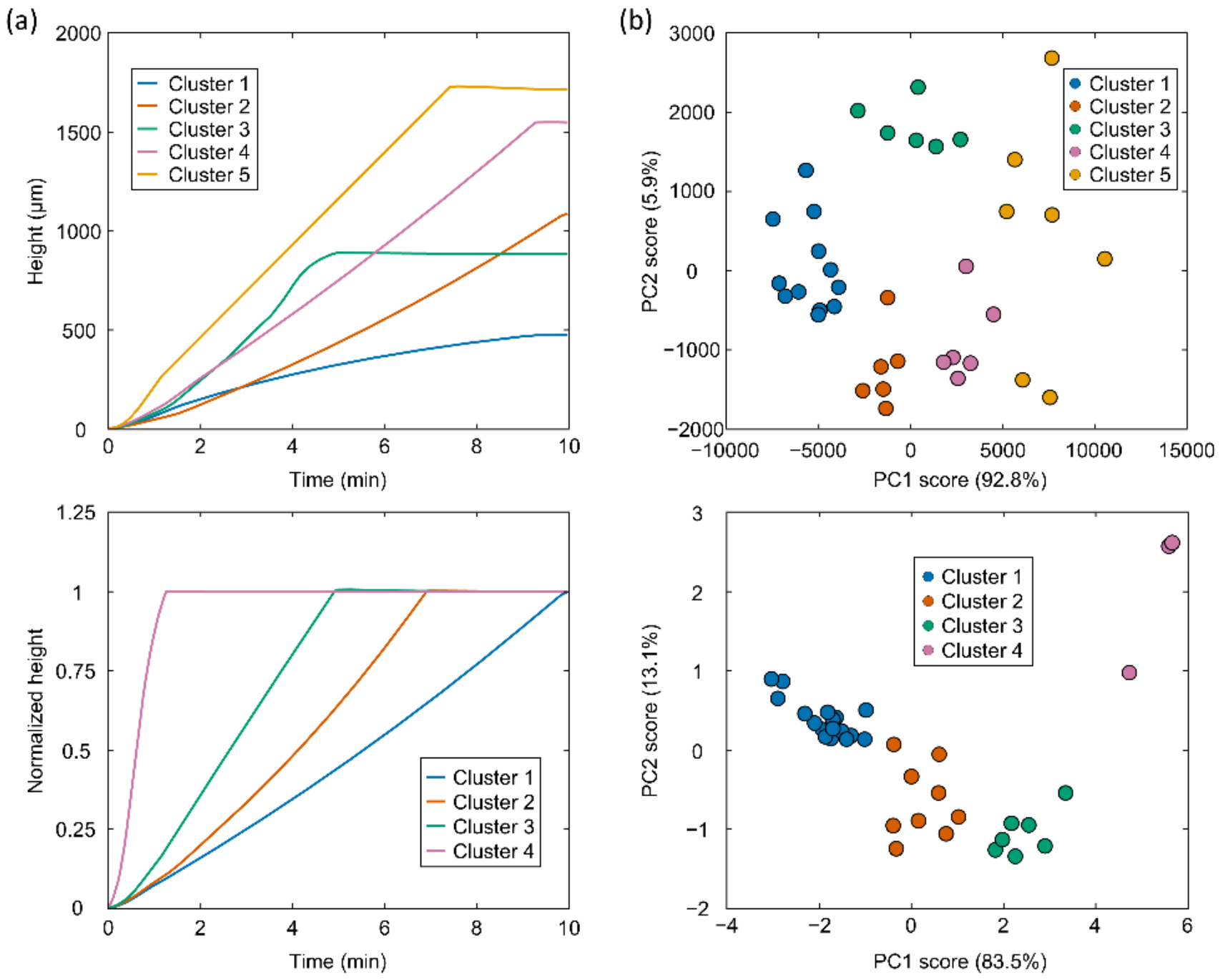

**Fig. 3.** Analysis of in-situ VACNT growth profiles. (a) Representative raw growth profiles obtained by hierarchical clustering and (b) the corresponding PCA score plot. (c) Representative normalized

growth profiles obtained by hierarchical clustering and (d) the corresponding PCA score plot. Five and four representative growth patterns were identified for the raw and normalized profiles, respectively.

### 3.4 Prediction of VACNT properties from gas conditions

To determine what information the growth curves contribute, we first asked how much of this observed variation in the final forest properties can be explained by the gas-flow conditions alone. The gas conditions captured much of the variation in the CNT density and G/D ratio, but not the variation in the CNT yield, at both the high and low ends of the yield range (Fig. 4). We determined this by training machine-learning models that used only the gas flow conditions as input features and comparing the experimentally observed and predicted values for each property (see Experimental).

For CNT density, the predicted values followed the parity line over most of the measured range, from approximately 20 to 50 mg $cm^{-3}$ (Fig. 4b), but forests with densities above approximately 55 mg $cm^{-3}$ were underestimated. The predicted G/D ratios likewise agreed with the measured values between approximately 4 and 11, whereas the few forests with G/D ratios above approximately 12 were underestimated (Fig. 4c). This indicates that the gas conditions contain information relevant to the structural quality of the resulting CNTs.

For CNT yield (Fig. 4a), the predictions did not reproduce the spread of the measured values. Whereas the measured yields ranged from approximately 0.3 to 4.4 mg $cm^{-2}$, most predicted values fell between approximately 1.5 and 2.5 mg $cm^{-2}$. Consequently, forests with measured yields above approximately 3 mg $cm^{-2}$ lay below the parity line and were underestimated, whereas those below approximately 1 mg $cm^{-2}$ lay above it and were overestimated. This compression of the predictions toward the middle of the measured range indicates that the gas-condition-based models did not fully capture the variation in CNT yield under the present dataset and modeling conditions. Notably, the training predictions were compressed into a similar range, indicating that the gas-flow inputs could not reproduce the measured yields even for samples used to train the model. This is consistent with forests grown under similar gas conditions reaching different yields.

This contrast is consistent with the complex relationships between the individual gas flow conditions and forest properties shown in Fig. 2. The gas conditions define the nominal synthesis environment, but they do not fully describe the state of the growth process. Even if the gas conditions cannot explain the variation in yield, they may still explain the differences in how the forests grew (Fig. 3). We therefore next examined how well the gas conditions alone predict the growth characteristics extracted from the growth curves.

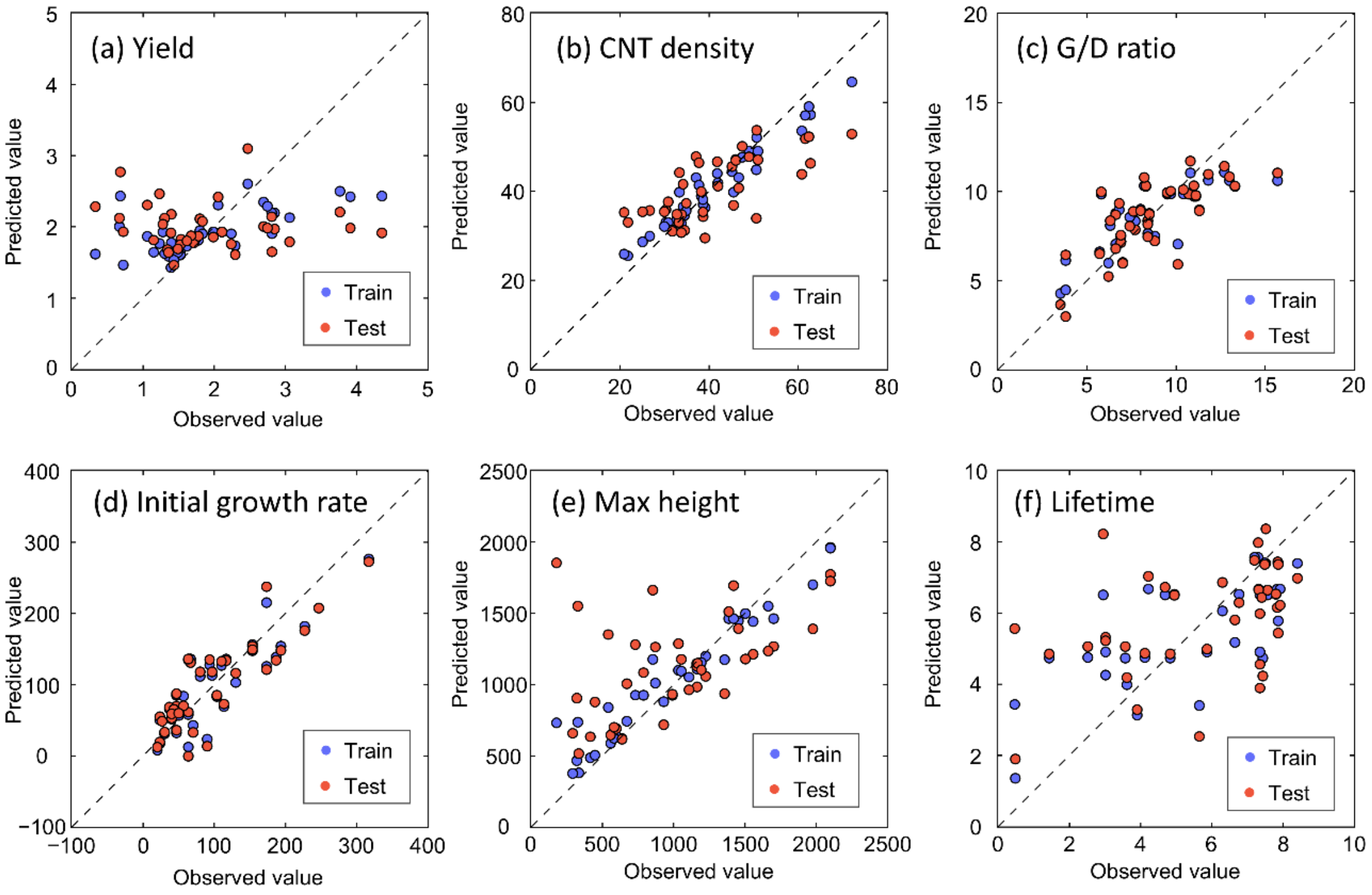

**Fig. 4.** Machine-learning prediction of VACNT forest properties and growth kinetics using only the gas flow conditions as input features. Experimentally observed and predicted values are compared for (a) CNT yield, (b) CNT density, (c) G/D ratio, (d) initial growth rate, (e) max height, and (f) growth lifetime. Blue points show predictions for the training data; red points show out-of-fold (OOF) predictions from 5-fold cross-validation, in which each sample was predicted by a model not trained on it. The dashed diagonal lines indicate ideal agreement between the observed and predicted values.

### 3.5 Prediction of VACNT growth characteristics from gas conditions

Forest height, and therefore yield, depends on both the growth speed and lifetime. The unexplained observed variation in yield may therefore originate in the growth kinetics. However, because *in-situ* measurements are difficult, the relationship between the gas conditions and the growth kinetics has rarely been examined. Here, we asked to what extent the gas conditions determine these kinetics. The gas conditions predicted the initial growth rate reasonably well but only partially captured the variation in maximum forest height and growth lifetime (Fig. 4d–f). We examined this relationship using models that, as in Fig. 4, used only the gas flow conditions as input features, to predict the initial growth rate, maximum forest height, and growth lifetime extracted from each growth curve.

For the initial growth rate, the predictions agreed with the measured values across the range of ~20–320 μm min$^{-1}$ (Fig. 4d), with larger scatter only for the slowest-growing forests, below ~ 70 μm min$^{-1}$. This scatter may originate from the difficulty in distinguishing the substrate surface from the forest uppermost surface when it is still very short.

In contrast, the deviations for maximum height were large and occurred across the entire measured range (Fig. 4e): several forests that reached only approximately 200–500 μm were predicted to reach 1000–1850 μm, whereas forests taller than approximately 1500 μm were generally underestimated. Growth lifetime showed the poorest agreement (Fig. 4f). For example, forests with measured lifetimes of approximately 7–7.5 min were assigned predicted lifetimes ranging from approximately 4 to 8 min, and most forests with lifetimes shorter than approximately 3 min were overestimated. Thus, the gas conditions alone could not reliably predict the persistence of the growth.

Together, these results suggest where the unexplained variation in yield originates. The gas conditions predicted the initial growth rate well (Fig. 4d) but not the growth lifetime (Fig. 4f). Because forest height is determined by both the initial growth rate and the growth lifetime [3,4], the large errors in the predicted maximum height (Fig. 4e) arise mainly from the poorly predicted lifetime rather than from the initial growth rate. Since yield is the product of forest height and density, and density was reasonably predicted (Fig. 4b), these height errors propagate directly to yield (Fig. 4a). In water-assisted CVD, growth termination is often governed by catalyst deactivation and gas delivery [3,4]. These results therefore suggest that the rate of catalyst deactivation varied among forests in ways not captured by the nominal gas conditions.

These findings highlight the distinction between the process inputs (the externally imposed gas conditions) and the observed process response (recorded by the *in-situ* growth curves). Together with the distinct growth patterns identified in Fig. 3, they suggest that direct observation of the growth process provides information complementary to the externally controlled synthesis conditions. If the growth curves record this variation in growth duration, adding them as model inputs would be expected to improve the prediction of properties that depend on it, most notably yield. We therefore next examined whether including the growth curves improves property prediction and how much each curve must be observed.

### 3.6 Contribution of time-resolved growth information to VACNT property prediction

If the unexplained variation in yield arises mainly from differences in growth lifetime (Fig. 4f), these differences should not be apparent at the onset of growth but should become visible in the growth curves as growth proceeds and slows. Yield prediction should therefore improve as more of the growth curve is observed. Density and G/D ratio are not directly recorded by the height curve, but they may still depend on the kinetics of the growth. For example, if CNT purity decreases during extended growth, the G/D ratio would decline with growth time, and the later portions of the curve would then contain information reflecting it [32].

To test these expectations, we combined the gas conditions with the growth curve recorded from the start of growth up to a given time (observation window). The endpoint of the observation window was progressively extended from 0 min, corresponding to the gas conditions alone, to 10 min, and the predictive performance of six regression methods was evaluated using the coefficient of determination, $R^2$ (see Experimental). Only yield benefited clearly, where its prediction improved as the observation window was extended, whereas density and G/D ratio showed no clear improvement (Fig. 5a–c).

For CNT yield, the predictive performance improved steadily as the observation window was extended (Fig. 5a). With the gas conditions alone, all models gave negative $R^2$ values (~ -1.3 to -0.2), meaning that their predictions were less accurate than simply using the mean yield. Including the first 2 minutes of the growth curve brought most models close to zero, and most became positive once the window reached 4 – 6 minutes. The best-performing models reached $R^2$ values of approximately 0.4–0.5 when the curve up to 8–10 min was included. Thus, information related to the final yield emerges mainly during the middle and later stages of growth. This is consistent with the differences in growth persistence and termination identified in Fig. 3. However, because the later observation windows also contain the forest height already reached, this improvement cannot yet be attributed to growth kinetics alone.

For CNT forest density, adding the growth curve did not improve prediction (Fig. 5b). Several models showed modest predictive performance using the gas conditions alone ($R^2 \approx 0.1$–$0.3$), but for most models, $R^2$ fell to near zero or below once the growth curve was added. Some models fluctuated strongly, particularly for intermediate observation windows (for one model, $R^2$ fell to approximately −0.65 at 4 min), whereas some models maintained $R^2$ values of approximately 0–0.2 throughout. This decline may reflect the larger number of input features relative to the dataset size, rather than an absence of relevant information in the growth curve. Nonetheless, including the growth curve did not improve density prediction beyond that achieved with the gas conditions alone.

For the G/D ratio, most models showed low but stable positive $R^2$ values (approximately 0.1–0.2) over nearly the entire range of observation windows (Fig. 5c). Extending the growth curve produced no clear improvement. Thus, neither the gas conditions nor the growth curve captured most of the variation in G/D ratio under the present dataset and modeling conditions. This suggests that the crystallinity and purity of the CNTs depend on factors not represented by either input. This result also suggests that, within the 10-min growth period, the G/D ratio did not change strongly with growth duration, or that any such change was too weak to detect at this level of predictability.

Overall, the value of the growth curve depended on the target property. The middle and later stages were informative for yield, whereas including growth curve information did not significantly improve the prediction of CNT forest density or G/D ratio. These differences indicate that the final VACNT properties reflect different aspects of the growth process and that the required duration of *in-situ* monitoring depends on the property to be predicted. Because yield depends largely on the attained forest height, we next examined whether the maximum height and growth lifetime could be predicted before the end of the growth period.

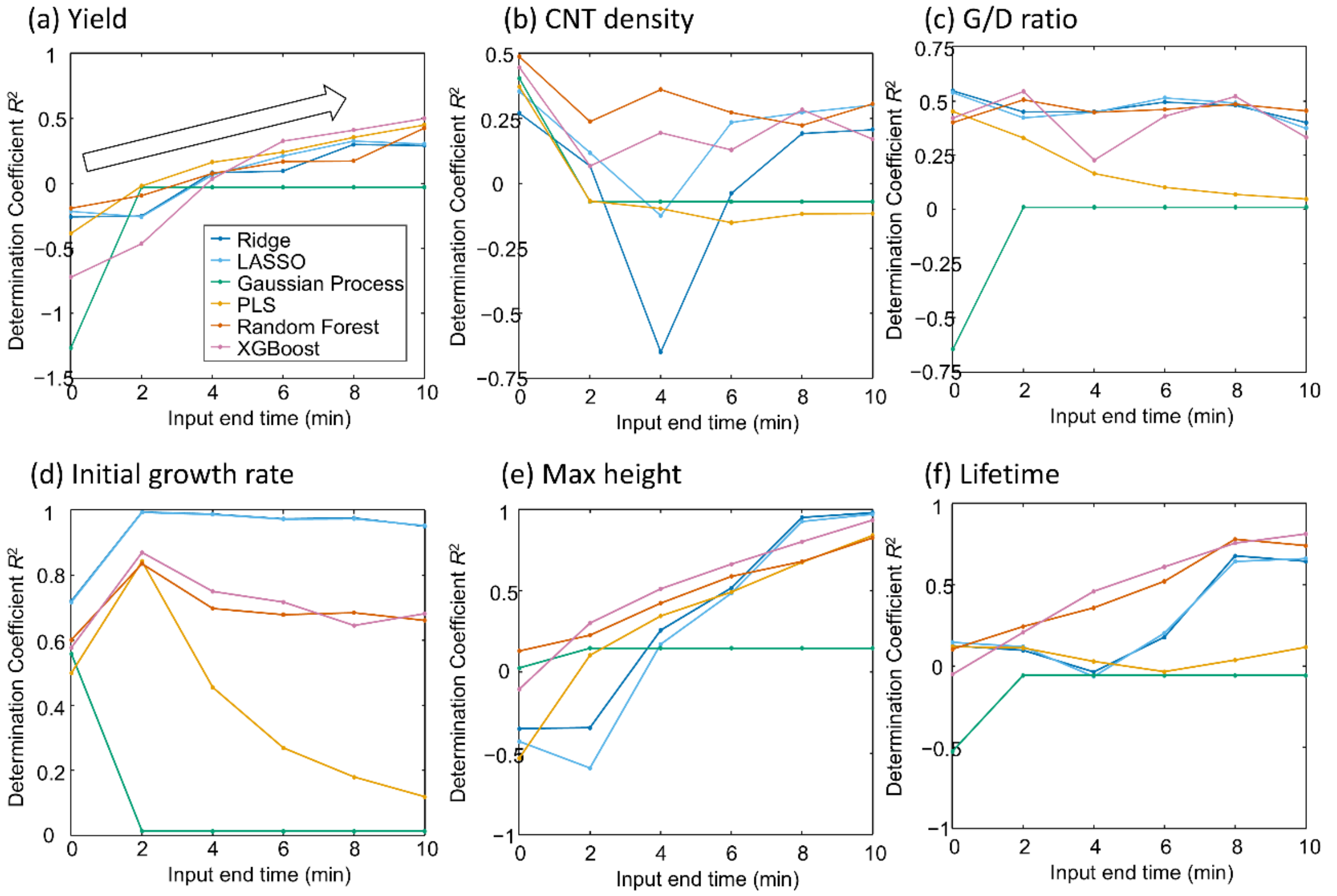

**Fig. 5.** Contribution of time-resolved VACNT growth information to the prediction of final forest properties and growth kinetics. The coefficient of determination ($R^2$) is plotted as a function of the endpoint of the in-situ height-profile observation window for the prediction of (a) yield, (b) CNT density, (c) G/D ratio, (d) initial growth rate, (e) maximum height and (f) growth lifetime. The models used the gas flow conditions together with partial height profiles acquired from the beginning of growth to the indicated observation-window endpoint. Results from Ridge, LASSO, Gaussian process regression, PLS regression, random forest, and XGBoost are shown. The machine-learning models used the gas flow conditions together with partial height profiles acquired from the beginning of growth to the indicated observation-window endpoint. An endpoint of 0 min represents prediction using only the gas conditions.

### 3.7 Time-dependent predictability of forest growth characteristics

As shown above, yield became predictable only once the middle and later stages of growth were observed (Fig. 5a–c). This is not simply a consequence of observing more of the growth. The first 2 min of the curve already contain the initial growth rate, which the gas conditions predicted well (Fig. 4d), yet including them left most yield models near $R^2 \approx 0$. Thus, the information relevant to yield lies not in how fast a forest begins to grow, but in how its growth progresses afterward. However, how predictable are the growth kinetics? Specifically, can the maximum height and the growth lifetime be anticipated before growth ends? Our results show that the initial growth rate was largely predictable from the gas conditions alone, whereas the maximum height and the growth lifetime could be estimated with moderate accuracy only from the middle stage of growth (4–6 min; Fig. 5d–f). This is reasonable because, by this stage, the extent of growth rate decay has been captured in the growth curve, and this decay reflects the balance between the growth conditions and the catalyst system [3,4]. To determine this, the same observation-window analysis used in Fig. 5a–c was applied to the initial growth rate, the maximum height, and the growth lifetime.

For the initial growth rate, the gas conditions alone already gave positive $R^2$ values (approximately 0.5–0.7; Fig. 5d), consistent with Fig. 4d. Ridge and LASSO regression reached $R^2$ values close to 1 once the first 2 min were included, which is expected because this window contains the 1-min height. Thus, the initial growth rate is largely set by the gas conditions.

In contrast, the maximum height depended strongly on the observation window (Fig. 5e). With the gas conditions alone, $R^2$ ranged from approximately −0.5 to 0.1, and including the first 2 min produced little improvement. Predictive performance rose steadily from 4 min onward. By 6 min, the best models (XGBoost and random forest) reached $R^2 \approx 0.6$–0.65, and Ridge and LASSO reached $R^2 \approx 0.95$ by 8–10 min, when most of the growth-rate decay had been observed.

Growth lifetime showed a similar trend, as expected from its relationship with the maximum height, with somewhat lower and more model-dependent predictability (Fig. 5f). With the gas conditions alone, $R^2$ values were low (approximately −0.5 to 0.1). XGBoost and random forest reached $R^2 \approx 0.5$–0.6 by 6 min, whereas Ridge and LASSO improved substantially only at 8 min, with the best models reaching approximately 0.75–0.8 at 8–10 min.

Taken together, the maximum height and growth lifetime could be moderately predicted by approximately 6 min, roughly halfway through the 10-min growth period and while the longer-lived forests were still growing. This timing parallels the yield results in Fig. 5a–c, where most models became positive only at 4–6 min, and is consistent with the interpretation that the unexplained variation in yield is linked to growth lifetime.

Among the final properties, yield was the only one for which incorporating the growth curve clearly improved predictive performance. However, the later observation windows captured both temporal growth behavior and attained forest height, preventing this analysis from distinguishing their respective contributions to yield prediction. We therefore examined the relative importance of measurements from different stages of growth in the yield model.

### 3.8 Temporal distribution of predictive information for CNT yield

The observation-window analysis identified when sufficient information for yield prediction had accumulated. However, because each window includes all preceding measurements, it could not reveal which portions of the growth curve the model relied on, or which gas variables contributed alongside the growth curve. To resolve these contributions, permutation feature importance was calculated for the XGBoost model constructed using the gas conditions and the full growth curve (Fig. 6). The importance of each feature was quantified as the increase in mean absolute error (MAE) caused by randomly permuting that feature with a larger increase indicating a greater contribution of the corresponding feature to the model prediction.

All growth curve measurements exhibited positive importance (Fig. 6a). The importance increased from approximately 0.035 mg cm$^{-2}$ at the onset of growth to approximately 0.1 mg cm$^{-2}$ within the first minute. It then rose gradually to approximately 0.14–0.15 mg cm$^{-2}$ between 6 and 8 min and, following a brief decrease near 9 min, reached a maximum of approximately 0.165 mg cm$^{-2}$ at 10 min. This temporal distribution is consistent with the observation-window analysis in Fig. 5a–c, in which the predictive performance for CNT yield improved as progressively longer portions of the growth curve were included. Notably, the early measurements also carried appreciable importance, although including them alone produced little improvement in yield prediction (Fig. 5a). This suggests that the early measurements contribute mainly in combination with the later ones, for example by providing a baseline (the initial growth rate) against which the subsequent slowing of growth can be gauged.

Among the gas variables, water-vapor supply and hydrogen supply exhibited comparable importance (approximately 0.05 mg cm$^{-2}$), whereas acetylene content contributed little (approximately 0.01 mg cm$^{-2}$; Fig. 6b). Water vapor is known to sustain catalyst activity [2], and $H_2$ can also influence the catalyst state. These observations suggest that the early measurements are informative primarily in combination with the later ones, likely because they establish the initial growth rate against which the subsequent growth-rate decay is quantified.

Two considerations limit the interpretation of these results. First, adjacent measurements along the growth curve are strongly correlated; their importance values are therefore not independent and cannot be compared directly with those of the individual gas variables. Second, the measurement at 10 min corresponds to the final forest height, which directly determines the yield together with the forest density. The maximum importance at 10 min is therefore likely to reflect the attained height rather than the growth kinetics alone. Consequently, permutation importance alone does not establish whether the contribution of the later measurements arises from the growth kinetics, the attained height, or a combination of both.

Among the final properties, yield was the only one for which incorporating the growth curve clearly improved predictive performance. We therefore examined the yield model in more detail. Overall, these results demonstrate that the *in-situ* growth curve provides predictive information for CNT yield beyond that represented by the gas conditions in the fitted XGBoost model, with the greatest contributions from the intermediate and later stages of growth. Identifying the specific kinetic descriptors that govern yield will require separating the contributions of the growth curve profile and the attained height, for example by comparing models based on the gas conditions and final height with those based on the full growth curve.

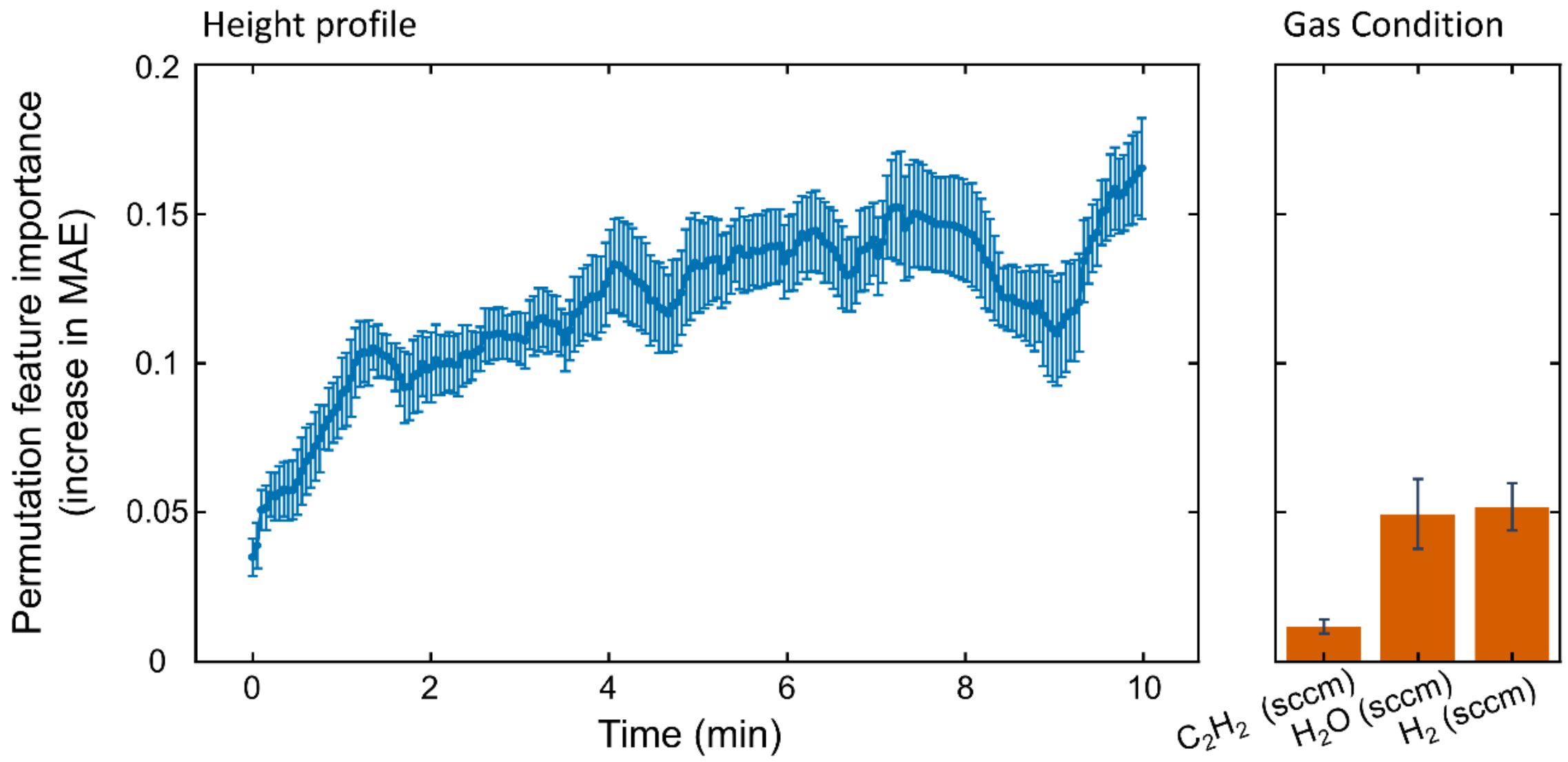

**Fig. 6.** Permutation feature importance of time-resolved growth and gas-flow variables for CNT yield prediction.

### 3.9 Implications for VACNT synthesis and autonomous experimentation

Taken together, these results show that the growth characteristics differ not in whether they can be predicted, but in when the necessary information becomes available. The initial growth rate is set by the synthesis conditions before growth begins, whereas the maximum height and the growth lifetime become predictable only as growth proceeds. Because the growth lifetime limits the height reached within the growth period and, through it, the yield, the variation in yield left unexplained by the gas conditions appears to originate in how catalyst activity evolves during growth, which the initial gas conditions do not capture. This variation should therefore be regarded not merely as experimental noise to be averaged out, but as a signature of the process state. This distinction implies that data-driven approaches that map synthesis conditions directly to final properties, as in most previous studies of CNT synthesis [21–24], are likely to be limited by the portion of the outcome that is determined during growth.

The benefit of *in-situ* monitoring is, however, limited by what the monitored signal records. The growth curve records how the height of the forest grows, which explains why it improved the prediction of yield but not of density or G/D ratio (Fig. 5b,c). Density depends on how closely the CNTs are packed, and the G/D ratio reflects their crystallinity and purity. While related, neither is directly recorded by forest height. Predicting these properties during growth will therefore require complementary *in-situ* signals, such as Raman spectroscopy, which has already been integrated into autonomous CNT synthesis [21]. Combining such signals with the observation-window analysis would extend decision points from the amount of growth to the structure of the grown material.

The observation-window formulation used here addresses this limitation by treating prediction as a question of timing. Rather than asking only whether a property can be predicted, it identifies when sufficient information becomes available during a process. In the present system, the maximum height and the growth lifetime could be estimated with moderate accuracy by approximately midway through the arbitrarily set growth period. This defines a decision point within a single batch. In autonomous experimentation, such a decision point could be used first to terminate unpromising batches early, reducing the time spent on poor conditions and maintaining quality of the grown material. In the longer term, it could be used to redirect the synthesis conditions within a batch while the catalyst remains active. Candidate variables include the water-vapor and $H_2$ supplies, which showed the greatest importance among the gas variables for yield prediction (Fig. 6b). Realizing such in-batch adjustment will require models trained on experiments in which the conditions are varied during growth, because the present models were trained only on growth under fixed conditions and may not remain valid once the trajectory is perturbed. More generally, the approach is not specific to CNT synthesis. Other synthesis processes monitored *in-situ* could be analyzed in the same way to determine which properties are set by the inputs and when the remaining ones become predictable.

## 4. Conclusions

Synthesis conditions are commonly treated as a sufficient description of a CVD growth, with the remaining variation in outcome attributed to experimental noise. By integrating *in-situ* growth curves with gas-flow conditions, we show that this description is incomplete for CNT forests. The gas conditions set how growth begins but not how long it persists, and this unpredicted growth lifetime appears to account for much of the yield variation they leave unexplained. This variation is therefore not merely noise but a signature of the evolving catalyst state and gas delivery, which becomes visible only as growth proceeds. The relevant question is thus not only whether a growth outcome can be predicted, but when. Here, the maximum height and lifetime could be estimated with moderate accuracy midway through growth, while the longer-lived forests were still growing, which defines a decision point within a single batch. This perspective shifts the role of *in-situ* monitoring from post hoc characterization of growth kinetics toward data-driven recognition of an evolving process state. For autonomous synthesis, the next step is therefore not only to search more efficiently for optimal conditions, but also to recognize during growth when a batch is diverging from its expected course, and to respond while the catalyst remains active.

## Author Contributions

Conceptualization: S.M., Methodology: S.M., Software: S.M., Formal Analysis: S.M., Investigation: S.M., T.T., S.S. D. N. F., Visualization: S.M., Writing – Original Draft: S.M. Writing - Review & Editing: T.T., D.N.F.

**Conflicts of interest**

The author declares having no conflicts of interest or competing interests.

**Data availability**

The model prediction accuracy is summarized in Table S1 of the Supporting Information. Additional data supporting the findings of this study are available from the corresponding author upon reasonable request.

**Acknowledgements**

We gratefully acknowledge the support provided by the members of the Nanocarbon Material Research Institute in AIST.

**Table S1.** Prediction accuracy of machine learning models.

| Category of target property | Name of the target property | Category of input variables | Model | R2 (train) | R2 (test) | MAE (train) | MAE (test) | RMSE (train) | RMSE (test) |
|---|---|---|---|---|---|---|---|---|---|
| CNT forest property | Yield | Gas condition only | Ridge | 0.059 | -0.258 | 0.68 | 0.79 | 0.87 | 1.01 |
| CNT forest property | Yield | Gas condition only | LASSO | 0.059 | -0.215 | 0.68 | 0.79 | 0.87 | 0.99 |
| CNT forest property | Yield | Gas condition only | Gaussian Process | 0.986 | -1.266 | 0.06 | 1.04 | 0.11 | 1.35 |
| CNT forest property | Yield | Gas condition only | PLS | 0.049 | -0.384 | 0.69 | 0.83 | 0.88 | 1.06 |
| CNT forest property | Yield | Gas condition only | Random Forest | 0.348 | -0.191 | 0.53 | 0.76 | 0.73 | 0.98 |
| CNT forest property | Yield | Gas condition only | XGBoost | 0.405 | -0.721 | 0.49 | 0.84 | 0.69 | 1.18 |
| CNT forest property | Yield | Height profile only | Ridge | 0.569 | 0.339 | 0.47 | 0.55 | 0.59 | 0.73 |
| CNT forest property | Yield | Height profile only | LASSO | 0.569 | 0.339 | 0.47 | 0.55 | 0.59 | 0.73 |
| CNT forest property | Yield | Height profile only | Gaussian Process | 0.966 | -0.049 | 0.04 | 0.75 | 0.17 | 0.92 |
| CNT forest property | Yield | Height profile only | PLS | 0.504 | 0.451 | 0.53 | 0.56 | 0.63 | 0.67 |
| CNT forest property | Yield | Height profile only | Random Forest | 0.919 | 0.494 | 0.18 | 0.49 | 0.26 | 0.64 |
| CNT forest property | Yield | Height profile only | XGBoost | 0.936 | 0.388 | 0.15 | 0.55 | 0.23 | 0.70 |
| CNT forest property | Yield | Gas condition and height profile | Ridge | 0.547 | 0.292 | 0.47 | 0.57 | 0.60 | 0.76 |
| CNT forest property | Yield | Gas condition and height profile | LASSO | 0.569 | 0.305 | 0.47 | 0.56 | 0.59 | 0.75 |
| CNT forest property | Yield | Gas condition and height profile | Gaussian Process | 1.000 | -0.029 | 0.00 | 0.74 | 0.00 | 0.91 |
| CNT forest property | Yield | Gas condition and height profile | PLS | 0.504 | 0.451 | 0.53 | 0.56 | 0.63 | 0.67 |
| CNT forest property | Yield | Gas condition and height profile | Random Forest | 0.791 | 0.426 | 0.31 | 0.51 | 0.41 | 0.68 |
| CNT forest property | Yield | Gas condition and height profile | XGBoost | 0.989 | 0.501 | 0.07 | 0.47 | 0.09 | 0.64 |
| CNT forest property | CNT density | Gas condition only | Ridge | 0.517 | 0.270 | 6.46 | 8.05 | 8.23 | 10.11 |
| CNT forest property | CNT density | Gas condition only | LASSO | 0.546 | 0.354 | 6.41 | 7.58 | 7.98 | 9.52 |
| CNT forest property | CNT density | Gas condition only | Gaussian Process | 0.973 | 0.404 | 0.93 | 7.09 | 1.94 | 9.14 |
| CNT forest property | CNT density | Gas condition only | PLS | 0.458 | 0.372 | 6.90 | 7.36 | 8.72 | 9.38 |
| CNT forest property | CNT density | Gas condition only | Random Forest | 0.915 | 0.488 | 2.69 | 6.68 | 3.46 | 8.47 |
| CNT forest property | CNT density | Gas condition only | XGBoost | 0.867 | 0.446 | 3.13 | 7.21 | 4.32 | 8.81 |
| CNT forest property | CNT density | Height profile only | Ridge | 0.137 | -0.197 | 8.62 | 10.47 | 11.00 | 12.96 |
| CNT forest property | CNT density | Height profile only | LASSO | 0.137 | -0.197 | 8.62 | 10.47 | 11.00 | 12.96 |
| CNT forest property | CNT density | Height profile only | Gaussian Process | 1.000 | -0.068 | 0.06 | 9.83 | 0.21 | 12.24 |
| CNT forest property | CNT density | Height profile only | PLS | 0.018 | -0.116 | 9.18 | 10.02 | 11.73 | 12.51 |
| CNT forest property | CNT density | Height profile only | Random Forest | 0.800 | -0.147 | 4.31 | 10.01 | 5.29 | 12.68 |
| CNT forest property | CNT density | Height profile only | XGBoost | 0.759 | -0.277 | 4.21 | 10.10 | 5.81 | 13.38 |
| CNT forest property | CNT density | Gas condition and height profile | Ridge | 0.561 | 0.207 | 6.43 | 8.64 | 7.85 | 10.54 |
| CNT forest property | CNT density | Gas condition and height profile | LASSO | 0.590 | 0.302 | 6.30 | 8.18 | 7.59 | 9.90 |
| CNT forest property | CNT density | Gas condition and height profile | Gaussian Process | 1.000 | -0.070 | 0.01 | 9.88 | 0.01 | 12.25 |
| CNT forest property | CNT density | Gas condition and height profile | PLS | 0.018 | -0.116 | 9.18 | 10.02 | 11.73 | 12.51 |
| CNT forest property | CNT density | Gas condition and height profile | Random Forest | 0.907 | 0.307 | 2.83 | 7.94 | 3.62 | 9.86 |
| CNT forest property | CNT density | Gas condition and height profile | XGBoost | 0.955 | 0.228 | 2.15 | 8.49 | 2.52 | 10.40 |
| CNT forest property | G/D ratio | Gas condition only | Ridge | 0.577 | 0.548 | 1.36 | 1.38 | 1.76 | 1.82 |
| CNT forest property | G/D ratio | Gas condition only | LASSO | 0.595 | 0.540 | 1.32 | 1.39 | 1.72 | 1.83 |
| CNT forest property | G/D ratio | Gas condition only | Gaussian Process | 0.976 | -0.645 | 0.23 | 2.59 | 0.42 | 3.47 |
| CNT forest property | G/D ratio | Gas condition only | PLS | 0.540 | 0.452 | 1.42 | 1.54 | 1.83 | 2.00 |
| CNT forest property | G/D ratio | Gas condition only | Random Forest | 0.684 | 0.402 | 1.19 | 1.71 | 1.52 | 2.09 |
| CNT forest property | G/D ratio | Gas condition only | XGBoost | 0.736 | 0.422 | 1.00 | 1.58 | 1.39 | 2.06 |
| CNT forest property | G/D ratio | Height profile only | Ridge | 0.312 | 0.113 | 1.70 | 1.98 | 2.24 | 2.55 |
| CNT forest property | G/D ratio | Height profile only | LASSO | 0.312 | 0.113 | 1.70 | 1.98 | 2.24 | 2.55 |
| CNT forest property | G/D ratio | Height profile only | Gaussian Process | 0.999 | 0.012 | 0.03 | 2.10 | 0.10 | 2.69 |
| CNT forest property | G/D ratio | Height profile only | PLS | 0.114 | 0.048 | 1.90 | 1.96 | 2.55 | 2.64 |
| CNT forest property | G/D ratio | Height profile only | Random Forest | 0.804 | 0.317 | 0.89 | 1.62 | 1.20 | 2.23 |
| CNT forest property | G/D ratio | Height profile only | XGBoost | 0.965 | 0.227 | 0.39 | 1.76 | 0.50 | 2.38 |
| CNT forest property | G/D ratio | Gas condition and height profile | Ridge | 0.591 | 0.401 | 1.37 | 1.63 | 1.73 | 2.09 |
| CNT forest property | G/D ratio | Gas condition and height profile | LASSO | 0.610 | 0.385 | 1.33 | 1.60 | 1.69 | 2.12 |
| CNT forest property | G/D ratio | Gas condition and height profile | Gaussian Process | 1.000 | 0.010 | 0.00 | 2.10 | 0.00 | 2.69 |
| CNT forest property | G/D ratio | Gas condition and height profile | PLS | 0.114 | 0.048 | 1.90 | 1.96 | 2.55 | 2.64 |
| CNT forest property | G/D ratio | Gas condition and height profile | Random Forest | 0.849 | 0.455 | 0.76 | 1.48 | 1.05 | 2.00 |
| CNT forest property | G/D ratio | Gas condition and height profile | XGBoost | 0.974 | 0.333 | 0.35 | 1.69 | 0.44 | 2.21 |
| In-situ | Initial growth rate | Gas condition only | Ridge | 0.766 | 0.721 | 28.31 | 30.25 | 33.66 | 36.77 |
| In-situ | Initial growth rate | Gas condition only | LASSO | 0.766 | 0.718 | 28.29 | 30.04 | 33.66 | 36.98 |
| In-situ | Initial growth rate | Gas condition only | Gaussian Process | 0.990 | 0.558 | 3.25 | 34.88 | 7.00 | 46.25 |
| In-situ | Initial growth rate | Gas condition only | PLS | 0.584 | 0.499 | 35.49 | 37.89 | 44.86 | 49.25 |
| In-situ | Initial growth rate | Gas condition only | Random Forest | 0.947 | 0.601 | 12.70 | 33.91 | 15.95 | 43.94 |
| In-situ | Initial growth rate | Gas condition only | XGBoost | 0.877 | 0.578 | 19.78 | 35.67 | 24.44 | 45.23 |
| In-situ | Initial growth rate | Height profile only | Ridge | 0.991 | 0.978 | 4.67 | 7.13 | 6.50 | 10.30 |
| In-situ | Initial growth rate | Height profile only | LASSO | 0.991 | 0.978 | 4.67 | 7.13 | 6.50 | 10.30 |
| In-situ | Initial growth rate | Height profile only | Gaussian Process | 1.000 | 0.015 | 0.05 | 53.00 | 0.07 | 69.07 |
| In-situ | Initial growth rate | Height profile only | PLS | 0.213 | 0.118 | 42.33 | 45.40 | 61.73 | 65.35 |
| In-situ | Initial growth rate | Height profile only | Random Forest | 0.956 | 0.588 | 9.72 | 30.89 | 14.54 | 44.68 |
| In-situ | Initial growth rate | Height profile only | XGBoost | 0.995 | 0.628 | 3.62 | 29.39 | 4.79 | 42.46 |
| In-situ | Initial growth rate | Gas condition and height profile | Ridge | 0.992 | 0.976 | 4.56 | 7.33 | 6.38 | 10.71 |
| In-situ | Initial growth rate | Gas condition and height profile | LASSO | 0.992 | 0.973 | 4.42 | 7.91 | 6.30 | 11.41 |
| In-situ | Initial growth rate | Gas condition and height profile | Gaussian Process | 1.000 | 0.012 | 0.05 | 53.79 | 0.07 | 69.16 |
| In-situ | Initial growth rate | Gas condition and height profile | PLS | 0.213 | 0.118 | 42.33 | 45.40 | 61.73 | 65.35 |
| In-situ | Initial growth rate | Gas condition and height profile | Random Forest | 0.957 | 0.649 | 9.74 | 30.52 | 14.42 | 41.23 |
| In-situ | Initial growth rate | Gas condition and height profile | XGBoost | 0.991 | 0.658 | 4.94 | 28.60 | 6.71 | 40.67 |
| In-situ | Max height | Gas condition only | Ridge | 0.119 | -0.349 | 392.28 | 461.92 | 482.66 | 597.29 |
| In-situ | Max height | Gas condition only | LASSO | 0.131 | -0.428 | 381.60 | 468.35 | 479.37 | 614.42 |
| In-situ | Max height | Gas condition only | Gaussian Process | 0.990 | 0.025 | 25.03 | 368.32 | 52.07 | 507.71 |
| In-situ | Max height | Gas condition only | PLS | 0.120 | -0.530 | 385.64 | 499.36 | 482.31 | 636.10 |
| In-situ | Max height | Gas condition only | Random Forest | 0.895 | 0.130 | 116.53 | 346.88 | 166.64 | 479.52 |
| In-situ | Max height | Gas condition only | XGBoost | 0.868 | -0.105 | 134.10 | 379.26 | 186.98 | 540.61 |
| In-situ | Max height | Height profile only | Ridge | 0.985 | 0.979 | 47.23 | 56.36 | 62.76 | 75.21 |
| In-situ | Max height | Height profile only | LASSO | 0.985 | 0.979 | 47.23 | 56.36 | 62.76 | 75.21 |
| In-situ | Max height | Height profile only | Gaussian Process | 1.000 | 0.167 | 0.40 | 388.43 | 0.47 | 469.35 |
| In-situ | Max height | Height profile only | PLS | 0.870 | 0.842 | 153.16 | 170.39 | 185.59 | 204.44 |
| In-situ | Max height | Height profile only | Random Forest | 0.987 | 0.892 | 47.25 | 132.13 | 57.95 | 169.07 |
| In-situ | Max height | Height profile only | XGBoost | 0.990 | 0.933 | 42.06 | 100.65 | 52.07 | 132.95 |
| In-situ | Max height | Gas condition and height profile | Ridge | 0.988 | 0.981 | 41.13 | 51.16 | 56.81 | 70.13 |
| In-situ | Max height | Gas condition and height profile | LASSO | 0.993 | 0.972 | 33.05 | 62.84 | 44.15 | 85.98 |
| In-situ | Max height | Gas condition and height profile | Gaussian Process | 1.000 | 0.148 | 0.40 | 404.73 | 0.47 | 474.75 |
| In-situ | Max height | Gas condition and height profile | PLS | 0.870 | 0.842 | 153.16 | 170.39 | 185.59 | 204.44 |
| In-situ | Max height | Gas condition and height profile | Random Forest | 0.975 | 0.826 | 60.70 | 159.19 | 81.10 | 214.19 |
| In-situ | Max height | Gas condition and height profile | XGBoost | 0.993 | 0.936 | 31.04 | 98.31 | 41.53 | 130.55 |
| In-situ | Lifetime | Gas condition only | Ridge | 0.497 | 0.124 | 1.29 | 1.68 | 1.60 | 2.11 |
| In-situ | Lifetime | Gas condition only | LASSO | 0.489 | 0.146 | 1.31 | 1.65 | 1.61 | 2.08 |
| In-situ | Lifetime | Gas condition only | Gaussian Process | 0.947 | -0.526 | 0.26 | 2.19 | 0.52 | 2.78 |
| In-situ | Lifetime | Gas condition only | PLS | 0.381 | 0.119 | 1.46 | 1.71 | 1.77 | 2.12 |
| In-situ | Lifetime | Gas condition only | Random Forest | 0.725 | 0.103 | 0.98 | 1.80 | 1.18 | 2.13 |
| In-situ | Lifetime | Gas condition only | XGBoost | 0.784 | -0.051 | 0.78 | 1.83 | 1.05 | 2.31 |
| In-situ | Lifetime | Height profile only | Ridge | 0.610 | 0.397 | 1.08 | 1.36 | 1.41 | 1.75 |
| In-situ | Lifetime | Height profile only | LASSO | 0.610 | 0.398 | 1.08 | 1.36 | 1.41 | 1.75 |
| In-situ | Lifetime | Height profile only | Gaussian Process | 1.000 | -0.055 | 0.00 | 1.98 | 0.00 | 2.32 |
| In-situ | Lifetime | Height profile only | PLS | 0.464 | 0.116 | 1.38 | 1.76 | 1.65 | 2.12 |
| In-situ | Lifetime | Height profile only | Random Forest | 0.975 | 0.821 | 0.23 | 0.67 | 0.36 | 0.96 |
| In-situ | Lifetime | Height profile only | XGBoost | 0.992 | 0.799 | 0.14 | 0.73 | 0.20 | 1.01 |
| In-situ | Lifetime | Gas condition and height profile | Ridge | 0.860 | 0.644 | 0.69 | 1.03 | 0.84 | 1.34 |
| In-situ | Lifetime | Gas condition and height profile | LASSO | 0.860 | 0.662 | 0.69 | 1.01 | 0.84 | 1.31 |
| In-situ | Lifetime | Gas condition and height profile | Gaussian Process | 1.000 | -0.058 | 0.00 | 2.00 | 0.00 | 2.32 |
| In-situ | Lifetime | Gas condition and height profile | PLS | 0.464 | 0.116 | 1.38 | 1.76 | 1.65 | 2.12 |
| In-situ | Lifetime | Gas condition and height profile | Random Forest | 0.964 | 0.713 | 0.30 | 0.88 | 0.43 | 1.21 |
| In-situ | Lifetime | Gas condition and height profile | XGBoost | 0.988 | 0.783 | 0.17 | 0.74 | 0.24 | 1.05 |